\documentstyle[prl,eqsecnum,aps]{revtex}

\begin{document}
\author{Jian-Qi Shen\footnote{E-mail address: jqshen@coer.zju.edu.cn}, Yi Jin \& Long Chen}
\address{Centre for Optical
and Electromagnetic Research, State Key Laboratory of Modern
Optical Instrumentation \\Zhejiang University, Hangzhou 310027, P.
R. China}
\date{\today }
\title{Preliminary Preparations: Scattering Problem of Double-layered Sphere \\Containing Left-handed Media\footnote{This paper will be submitted nowhere else for the publication, just uploaded at the e-print archives.}}
\maketitle

\begin{abstract}
In this paper, we present Mie coefficients of double-layered
sphere and consider the scattering problem, including the topics
on field distribution, electromagnetic cross section, extinction
spectra as well as some potential peculiar properties arising from
the presence of left-handed media. The formulation presented here
can be easily generalized to cases of multiple-layered spheres.
\\


\\ \\
\end{abstract}
\section{Introduction}
More recently, a kind of artificial composite media ( the
so-called left-handed media ) having a frequency band where the
effective permittivity ( $\varepsilon$ ) and the effective
permeability ( $\mu$ ) are simultaneously negative attracts
attention of many researchers in various fields such as materials
science, condensed matter physics, optics and
electromagnetism\cite{Smith,Klimov,Pendry3,Shelby,Ziolkowski2}.
Veselago first considered this peculiar medium and showed that it
possesses a negative index of refraction\cite{Veselago}. It
follows from the Maxwell's equations that in this medium the
Poynting vector and wave vector of electromagnetic wave would be
antiparallel, {\it i. e.}, the vector {\bf {k}}, the electric
field {\bf {E}} and the magnetic field {\bf {H}} form a
left-handed system; thus Veselago referred to such materials as
``left-handed'', and correspondingly, the ordinary medium in which
{\bf {k}}, {\bf {E}} and {\bf {H}} form a right-handed system may
be termed the ``right-handed'' medium. Other authors call this
class of materials ``negative-index media (NIM)'', ``double
negative media (DNM)''\cite{Ziolkowski2} and Veselago's media. It
is readily verified that in such media having both $\varepsilon$
and $\mu$ negative, there exist a number of peculiar
electromagnetic properties, for instance, many dramatically
different propagation characteristics stem from the sign change of
the group velocity, including reversals of both the Doppler shift
and the Cherenkov radiation, anomalous refraction, modified
spontaneous emission rates and even reversals of radiation
pressure to radiation tension\cite{Klimov}. In experiments, this
artificial negative electric permittivity media may be obtained by
using the array of long metallic wires (ALMWs), which simulates
the plasma behavior at microwave frequencies, and the artificial
negative magnetic permeability media may be built up by using
small resonant metallic particles, {\it e. g.}, the split ring
resonators (SRRs), with very high magnetic
polarizability\cite{Pendry3,Pendry1,Pendry2,Maslovski}.

The extinction properties of a sphere ( single-layered ) with
negative permittivity and permeability is investigated by Ruppin.
Since recently Wang and Asher developed a novel method to
fabricate nanocomposite ${\rm SiO}_{2}$ spheres ( $\sim 100$ nm )
containing homogeneously dispersed Ag quantum dots ( $2\sim 5$ nm
)\cite{Wang}, which may has potential applications to design and
fabrication of photonic crystals, it is believed that the
absorption and transmittance of double-layered sphere deserves
consideration. In this paper, both the Mie coefficients and the
electromagnetic field distributions in the two-layered sphere
 containing left-handed media irradiated by a planar
 electromagnetic wave are presented.
\section{Methods}
It is well known that in the absence of electromagnetic sources,
the characteristic vectors such as ${\bf E}$, ${\bf B}$, ${\bf
D}$, ${\bf H}$ and Hertz vector in the isotropic homogeneous media
agree with the same differential equation
\begin{equation}
\nabla\nabla\cdot{\bf C}-\nabla\times\nabla\times{\bf C}+k^{2}{\bf
C}=0.
\end{equation}
The three independent vector solutions to the above equation is
\cite{Stratton}
\begin{equation}
{\bf L}=\nabla\psi,   \quad   {\bf M}=\nabla\times{\bf a}\psi,
\quad             {\bf N}=\frac{1}{k}\nabla\times{\bf M}
\end{equation}
with ${\bf a}$ being a constant vector, where the scalar function
$\psi$ satisfies $\nabla^{2}\psi+k^{2}\psi=0$. It is verified that
the vector solutions ${\bf M}$, ${\bf N}$ and ${\bf L}$ possess
the following mathematical properties
\begin{equation}
{\bf M}={\bf L}\times{\bf a}=\frac{1}{k}\nabla\times{\bf N}, \quad
{\bf L}\cdot{\bf M}=0,    \quad   \nabla\times{\bf L}=0,    \quad
\nabla\cdot{\bf L}=\nabla^{2}\psi=-k^{2}\psi, \quad
\nabla\cdot{\bf M}=0,  \quad       \nabla\cdot{\bf N}=0.
\end{equation}
Set ${\bf M}={\bf m}\exp(-i\omega t)$ and ${\bf N}={\bf
n}\exp(-i\omega t)$, the vector wave functions ${\bf M}$ and ${\bf
N}$ can be expressed in terms of the following spherical vector
wave functions\cite{Stratton}
\begin{eqnarray}
{\bf m}_{^{e}_{o}mn}&=&\mp\frac{m}{\sin
\theta}z_{n}(kr)P^{m}_{n}(\cos \theta){^{\sin}_{\cos}}m\phi{\bf
i}_{2}-z_{n}(kr)\frac{\partial P^{m}_{n}(\cos \theta)}{\partial
\theta}{^{\cos}_{\sin}}m\phi{\bf i}_{3},    \nonumber   \\
{\bf n}_{^{e}_{o}mn}&=&\frac{n(n+1)}{kr}z_{n}(kr)P^{m}_{n}(\cos
\theta){^{\cos}_{\sin}}m\phi{\bf
i}_{1}+\frac{1}{kr}\frac{\partial}{\partial
r}[rz_{n}(kr)]\frac{\partial}{\partial \theta}P^{m}_{n}(\cos
\theta){^{\cos}_{\sin}}m\phi{\bf i}_{2}
                                       \nonumber   \\
&\mp&\frac{m}{kr\sin \theta }\frac{\partial}{\partial
r}[rz_{n}(kr)]P^{m}_{n}(\cos \theta){^{\sin}_{\cos}}m\phi{\bf
i}_{3}.
\end{eqnarray}
In what follows we treat the Mie coefficients of double-layered
sphere irradiated by a plane wave.

\subsection{Definitions}
We consider a double-layered sphere with interior radius $a_{1}$
and external radius $a_{2}$ having relative permittivity
(permeability) $\epsilon_{1}$ ($\mu_{1}$) and $\epsilon_{2}$ (
$\mu_{2}$), respectively, placed in a medium having the relative
permittivity $\epsilon_{0}$ and permeability $\mu_{0}$. Suppose
that the double-layered sphere irradiated by the following plane
wave with the electric amplitude $E_{0}$ along the \^{z}-direction
of Cartesian coordinate system\cite{Stratton}
\begin{eqnarray}
{\bf E}_{\rm i}&=&{\bf a}_{x}E_{0}\exp(ik_{0}z-i\omega
t)=E_{0}\exp(-i\omega
t)\sum^{\infty}_{n=1}i^{n}\frac{2n+1}{n(n+1)}\left({\bf
m}^{(1)}_{o1n}-i{\bf n}^{(1)}_{e1n}\right),  \nonumber \\
{\bf H}_{\rm i}&=&{\bf
a}_{y}\frac{k_{0}}{\mu_{0}\omega}E_{0}\exp(ik_{0}z-i\omega
t)=-\frac{k_{0}}{\mu_{0}\omega}E_{0}\exp(-i\omega
t)\sum^{\infty}_{n=1}i^{n}\frac{2n+1}{n(n+1)}\left({\bf
m}^{(1)}_{e1n}+i{\bf n}^{(1)}_{o1n}\right),
\end{eqnarray}
where
\begin{eqnarray}
{\bf m}^{(1)}_{^{o}_{e}1n}&=&\pm
\frac{1}{\sin\theta}j_{n}(k_{0}r)P^{1}_{n}(\cos\theta){^{\cos}_{\sin}\phi}{\bf
i}_{2}-j_{n}(k_{0}r)\frac{\partial P^{1}_{n}}{\partial
\theta}{^{\sin}_{\cos}\phi}{\bf i}_{3},                     \nonumber \\
{\bf
n}^{(1)}_{^{o}_{e}1n}&=&\frac{n(n+1)}{k_{0}r}j_{n}(k_{0}r)P^{1}_{n}(\cos\theta){^{\sin}_{\cos}\phi}{\bf
i}_{1}+\frac{1}{k_{0}r}\left[k_{0}rj_{n}(k_{0}r)\right]'\frac{\partial
P^{1}_{n}}{\partial \theta}{^{\sin}_{\cos}\phi}{\bf i}_{2}\pm
\frac{1}{k_{0}r\sin\theta}\left[k_{0}rj_{n}(k_{0}r)\right]'P^{1}_{n}(\cos\theta){^{\cos}_{\sin}\phi}{\bf
i}_{3}                      \label{eq26}
\end{eqnarray}
with the primes denoting differentiation with respect to their
arguments and $k_{0}$ being
$\sqrt{\epsilon_{0}\mu_{0}}\frac{\omega}{c}$. In the region
$r<a_{1}$, the wave function is expanded as the following series
\begin{eqnarray}
{\bf E}_{\rm t}&=&E_{0}\exp(-i\omega
t)\sum^{\infty}_{n=1}i^{n}\frac{2n+1}{n(n+1)}\left(a^{\rm
t}_{n}{\bf
m}^{(1)}_{o1n}-ib^{\rm t}_{n}{\bf n}^{(1)}_{e1n}\right),                        \nonumber \\
{\bf H}_{\rm t}&=&-\frac{k_{1}}{\mu_{1}\omega}E_{0}\exp(-i\omega
t)\sum^{\infty}_{n=1}i^{n}\frac{2n+1}{n(n+1)}\left(b^{\rm
t}_{n}{\bf m}^{(1)}_{e1n}+ia^{\rm t}_{n}{\bf
n}^{(1)}_{o1n}\right).
\end{eqnarray}
Note that here the propagation constant in ${\bf m}^{(1)}_{o1n}$
and ${\bf n}^{(1)}_{e1n}$ which have been defined in (\ref{eq26})
is replaced with $k_{1}$, {\it i.e.},
$\sqrt{\epsilon_{1}\mu_{1}}\frac{\omega}{c}$.
 In the region $a_{1}<r<a_{2}$, one may expand the electromagnetic
 wave amplitude as
\begin{eqnarray}
{\bf E}_{\rm m}&=&E_{0}\exp(-i\omega
t)\sum^{\infty}_{n=1}i^{n}\frac{2n+1}{n(n+1)}\left(a^{\rm
m}_{n}{\bf m}^{(1)}_{o1n}-ib^{\rm m}_{n}{\bf
n}^{(1)}_{e1n}+a^{\bar{\rm m}}_{n}{\bf m}^{(3)}_{o1n}-ib^{\bar{\rm
m}}_{n}{\bf n}^{(3)}_{e1n}\right),
\nonumber \\
{\bf H}_{\rm m}&=&-\frac{k_{2}}{\mu_{2}\omega}E_{0}\exp(-i\omega
t)\sum^{\infty}_{n=1}i^{n}\frac{2n+1}{n(n+1)}\left(b^{\rm
m}_{n}{\bf m}^{(1)}_{e1n}+ia^{\rm m}_{n}{\bf
n}^{(1)}_{o1n}+b^{\bar{\rm m}}_{n}{\bf m}^{(3)}_{e1n}+ia^{\bar{\rm
m}}_{n}{\bf n}^{(3)}_{o1n}\right).
\end{eqnarray}
Note that here in order to obtain ${\bf m}^{(3)}_{e1n}$ and ${\bf
n}^{(3)}_{o1n}$, the spherical Bessel functions $j_{n}$ in the
spherical vector function ${\bf m}^{(1)}_{e1n}$ and ${\bf
n}^{(1)}_{o1n}$ is replaced by the Hankel functions $h^{(1)}_{n}$,
namely, the explicit expressions for ${\bf m}^{(3)}_{e1n}$ and
${\bf n}^{(3)}_{o1n}$ can also be obtained from (\ref{eq26}), so
long as we replace the spherical Bessel functions $j_{n}$ in
(\ref{eq26}) with the Hankel functions $h^{(1)}_{n}$. Apparently,
here the propagation constant in ${\bf m}^{(1)}_{o1n}$ and ${\bf
n}^{(1)}_{e1n}$ should be replaced with $k_{2}$, {\it i.e.},
$\sqrt{\epsilon_{2}\mu_{2}}\frac{\omega}{c}$.

In the region $r>a_{2}$ the reflected wave is written
\cite{Stratton}
\begin{eqnarray}
{\bf E}_{\rm r}&=&E_{0}\exp(-i\omega
t)\sum^{\infty}_{n=1}i^{n}\frac{2n+1}{n(n+1)}\left(a^{\rm
r}_{n}{\bf
m}^{(3)}_{o1n}-ib^{\rm r}_{n}{\bf n}^{(3)}_{e1n}\right),                        \nonumber \\
{\bf H}_{\rm r}&=&-\frac{k_{0}}{\mu_{0}\omega}E_{0}\exp(-i\omega
t)\sum^{\infty}_{n=1}i^{n}\frac{2n+1}{n(n+1)}\left(b^{\rm
r}_{n}{\bf m}^{(3)}_{e1n}+ia^{\rm r}_{n}{\bf
n}^{(3)}_{o1n}\right),
\end{eqnarray}
where the propagation vector in ${\bf m}^{(3)}_{e1n}$ and ${\bf
n}^{(3)}_{o1n}$ is replaced by $k_{0}$, {\it i.e.},
$\sqrt{\epsilon_{0}\mu_{0}}\frac{\omega}{c}$.
\subsection{Boundary conditions}
At the boundary $r=a_{1}$, the boundary condition is given as
follows
\begin{equation}
{\bf i}_{1}\times{\bf E}_{\rm m}={\bf i}_{1}\times{\bf E}_{\rm t},
\quad {\bf i}_{1}\times{\bf H}_{\rm m}={\bf i}_{1}\times{\bf
H}_{\rm t},
\end{equation}
where the unit vectors of Cartesian coordinate system agree with
${\bf i}_{1}\times{\bf i}_{2}={\bf i}_{3}$, ${\bf i}_{2}\times{\bf
i}_{3}={\bf i}_{1}$, ${\bf i}_{3}\times{\bf i}_{1}={\bf i}_{2}$.

It follows from ${\bf i}_{1}\times{\bf E}_{\rm m}={\bf
i}_{1}\times{\bf E}_{\rm t}$ that
\begin{eqnarray}
a^{\rm m}_{n}j_{n}(N_{2}\rho_{1})&+&a^{\bar{\rm
m}}_{n}h^{(1)}_{n}(N_{2}\rho_{1})=a^{\rm
t}_{n}j_{n}(N_{1}\rho_{1}),
\nonumber  \\
N_{1}b^{\rm
m}_{n}\left[N_{2}\rho_{1}j_{n}(N_{2}\rho_{1})\right]'&+&N_{1}b^{\bar{\rm
m}}_{n}\left[N_{2}\rho_{1}h^{(1)}_{n}(N_{2}\rho_{1})\right]'=N_{2}b^{\rm
t}_{n}\left[N_{1}\rho_{1}j_{n}(N_{1}\rho_{1})\right]',
\end{eqnarray}
where $\rho_{1}=k_{0}a_{1}$, $N_{1}=\frac{k_{1}}{k_{0}}$.

In the similar manner, it follows from ${\bf i}_{1}\times{\bf
H}_{\rm m}={\bf i}_{1}\times{\bf H}_{\rm t}$ that
\begin{eqnarray}
\mu_{1}\left\{a^{\rm
m}_{n}\left[N_{2}\rho_{1}j_{n}(N_{2}\rho_{1})\right]'+a^{\bar{\rm
m}}_{n}\left[N_{2}\rho_{1}h^{(1)}_{n}(N_{2}\rho_{1})\right]'\right\}&=&\mu_{2}\left\{a^{\rm
t}_{n}\left[N_{1}\rho_{1}j_{n}(N_{1}\rho_{1})\right]'\right\},
\nonumber  \\
N_{2}\mu_{1}\left[b^{\rm m}_{n}j_{n}(N_{2}\rho_{1})+b^{\bar{\rm
m}}_{n}h^{(1)}_{n}(N_{2}\rho_{1})\right]&=&N_{1}\mu_{2}b^{\rm
t}_{n}j_{n}(N_{1}\rho_{1}).
\end{eqnarray}

At the boundary $r=a_{2}$, in the same fashion, it follows from
${\bf i}_{1}\times({\bf E}_{\rm i}+{\bf E}_{\rm r})={\bf
i}_{1}\times{\bf E}_{\rm m}$ that
\begin{eqnarray}
j_{n}(\rho_{2})+a^{\rm r}_{n}h^{(1)}_{n}(\rho_{2})&=&a^{\rm
m}_{n}j_{n}(N_{2}\rho_{2})+a^{\bar{\rm
m}}_{n}h^{(1)}_{n}(N_{2}\rho_{2}),
\nonumber  \\
N_{2}\left[\rho_{2}j_{n}(\rho_{2})\right]'+N_{2}b^{\rm
r}_{n}\left[\rho_{2}h^{(1)}_{n}(\rho_{2})\right]'&=&b^{\rm
m}_{n}\left[N_{2}\rho_{2}j_{n}(N_{2}\rho_{2})\right]'+b^{\bar{\rm
m}}_{n}\left[N_{2}\rho_{2}h^{(1)}_{n}(N_{2}\rho_{2})\right]',
\end{eqnarray}
where $\rho_{2}=k_{0}a_{2}$, $N_{2}=\frac{k_{2}}{k_{0}}$.

 In the meanwhile, it follows from
${\bf i}_{1}\times({\bf H}_{\rm i}+{\bf H}_{\rm r})={\bf
i}_{1}\times{\bf H}_{\rm m}$ that
\begin{eqnarray}
\mu_{2}\left\{a^{\rm
r}_{n}\left[\rho_{2}h^{(1)}_{n}(\rho_{2})\right]'+\left[\rho_{2}j_{n}(\rho_{2})\right]'\right\}&=&\mu_{0}\left\{a^{\rm
m}_{n}\left[N_{2}\rho_{2}j_{n}(N_{2}\rho_{2})\right]'+a^{\bar{\rm
m}}_{n}\left[N_{2}\rho_{2}h^{(1)}_{n}(N_{2}\rho_{2})\right]'\right\},
\nonumber  \\
\mu_{2}b^{\rm
r}_{n}h^{(1)}_{n}(\rho_{2})+\mu_{2}j_{n}(\rho_{2})&=&N_{2}\mu_{0}b^{\rm
m}_{n}j_{n}(N_{2}\rho_{2})+N_{2}\mu_{0}b^{\bar{\rm \rm
m}}_{n}h^{(1)}_{n}(N_{2}\rho_{2}).
\end{eqnarray}
\subsection{Calculation of Mie coefficients}
\subsubsection{Mie coefficient $a^{\rm r}_{n}$}
According to the above boundary conditions, one can arrive at the
following matrix equation
\begin{equation}
\left(\begin{array}{cccc}
-j_{n}(N_{1}\rho_{1})  & j_{n}(N_{2}\rho_{1}) & h^{(1)}_{n}(N_{2}\rho_{1})& 0 \\
-\mu_{2}\left[N_{1}\rho_{1}j_{n}(N_{1}\rho_{1})\right]' & \mu_{1}\left[N_{2}\rho_{1}j_{n}(N_{2}\rho_{1})\right]'& \mu_{1}\left[N_{2}\rho_{1}h^{(1)}_{n}(N_{2}\rho_{1})\right]' & 0  \\
 0 & j_{n}(N_{2}\rho_{2})&h^{(1)}_{n}(N_{2}\rho_{2}) &  -h^{(1)}_{n}(\rho_{2})   \\
 0 & \mu_{0}\left[N_{2}\rho_{2}j_{n}(N_{2}\rho_{2})\right]' &\left[N_{2}\rho_{2}h^{(1)}_{n}(N_{2}\rho_{2})\right]' &
 -\mu_{2}\left[\rho_{2}h^{(1)}_{n}(\rho_{2})\right]'
 \end{array}
 \right)\left(\begin{array}{cccc}
 a^{\rm t}_{n} \\
 a^{\rm m}_{n}\\
 a^{\bar{\rm m}}_{n}\\
  a^{\rm r}_{n}
 \end{array}
 \right)=\left(\begin{array}{cccc}
 0\\
 0\\
 j_{n}(\rho_{2})\\
 \mu_{2}\left[\rho_{2}j_{n}(\rho_{2})\right]'
 \end{array}
 \right).           \label{eq215}
\end{equation}

For convenience, the $4\times4$ matrix on the left handed side of
(\ref{eq215}) is denoted by
\begin{equation}
A=\left(\begin{array}{cccc}
A_{11} & A_{12} & A_{13}& 0 \\
A_{21} & A_{22}& A_{23}& 0  \\
 0 & A_{32} & A_{33} & A_{34}\\
 0&A_{42}&A_{43}&A_{44}
 \end{array}
 \right)
\end{equation}
whose determinant is
\begin{eqnarray}
{\rm det}
A&=&A_{11}A_{22}A_{33}A_{44}-A_{11}A_{22}A_{34}A_{43}-A_{11}A_{32}A_{23}A_{44}+A_{11}A_{42}A_{23}A_{34}
\nonumber  \\
&-&A_{21}A_{12}A_{33}A_{44}+A_{21}A_{12}A_{34}A_{43}+A_{21}A_{32}A_{13}A_{44}-A_{21}A_{42}A_{13}A_{34}.
\end{eqnarray}

So, the Mie coefficient $a^{\rm r}_{n}$ is of the form
\begin{equation} a^{\rm
r}_{n}=\frac{{\rm det} A_{\rm r}}{{\rm det} A}
\end{equation}
with
\begin{equation}
 A_{\rm r}=\left(\begin{array}{cccc}
A_{11} & A_{12} & A_{13}& 0 \\
A_{21} & A_{22}& A_{23}& 0  \\
 0 & A_{32} & A_{33} & j_{n}(\rho_{2})\\
 0&A_{42}&A_{43}&\mu_{2}\left[\rho_{2}j_{n}(\rho_{2})\right]'
 \end{array}
 \right)
\end{equation}
whose determinant is
\begin{eqnarray}
{\rm det} A_{\rm r
}&=&A_{11}A_{22}A_{33}\mu_{2}\left[\rho_{2}j_{n}(\rho_{2})\right]'-A_{11}A_{22}j_{n}(\rho_{2})A_{43}-A_{11}A_{32}A_{23}\mu_{2}\left[\rho_{2}j_{n}(\rho_{2})\right]'+A_{11}A_{42}A_{23}j_{n}(\rho_{2})
\nonumber   \\
&-&A_{21}A_{12}A_{33}\mu_{2}\left[\rho_{2}j_{n}(\rho_{2})\right]'+A_{21}A_{12}j_{n}(\rho_{2})A_{43}+A_{21}A_{32}A_{13}\mu_{2}\left[\rho_{2}j_{n}(\rho_{2})\right]'-A_{21}A_{42}A_{13}j_{n}(\rho_{2}).
\end{eqnarray}

\subsubsection{Mie coefficient $b^{\rm r}_{n}$}
According to the above boundary conditions, one can arrive at the
following matrix equation
 \begin{equation}
\left(\begin{array}{cccc}
-N_{2}\left[N_{1}\rho_{1}j_{n}(N_{1}\rho_{1})\right]' & N_{1}\left[N_{2}\rho_{1}j_{n}(N_{2}\rho_{1})\right]' & N_{1}\left[N_{2}\rho_{1}h^{(1)}_{n}(N_{2}\rho_{1})\right]' & 0 \\
-N_{1}\mu_{2}j_{n}(N_{1}\rho_{1}) & N_{2}\mu_{1}j_{n}(N_{2}\rho_{1}) & N_{2}\mu_{1}h^{(1)}_{n}(N_{2}\rho_{1})& 0  \\
 0 & \left[N_{2}\rho_{2}j_{n}(N_{2}\rho_{2})\right]' & \left[N_{2}\rho_{2}h^{(1)}_{n}(N_{2}\rho_{2})\right]'&
 -N_{2}\left[\rho_{2}h^{(1)}_{n}(\rho_{2})\right]'   \\
 0& N_{2}\mu_{0}j_{n}(N_{2}\rho_{2})& N_{2}\mu_{0}h^{(1)}_{n}(N_{2}\rho_{2})&
 -\mu_{2}h^{(1)}_{n}(\rho_{2})
 \end{array}
 \right)\left(\begin{array}{cccc}
b^{\rm t}_{n} \\
 b^{\rm m}_{n}\\
 b^{\bar{\rm m}}_{n}\\
  b^{\rm r}_{n}
 \end{array}
 \right)=\left(\begin{array}{cccc}
 0\\
 0\\
 N_{2}\left[\rho_{2}j_{n}(\rho_{2})\right]' \\
 \mu_{2}j_{n}(\rho_{2})
 \end{array}
 \right).                           \label{eq221}
\end{equation}

For convenience, the $4\times4$ matrix on the left handed side of
(\ref{eq221}) is denoted by
\begin{equation}
B=\left(\begin{array}{cccc}
B_{11} & B_{12} & B_{13}& 0 \\
B_{21} & B_{22}& B_{23}& 0  \\
 0 & B_{32} & B_{33} & B_{34}\\
 0&B_{42}&B_{43}&B_{44}
 \end{array}
 \right)
\end{equation}
whose determinant is
\begin{eqnarray}
{\rm det}
B&=&B_{11}B_{22}B_{33}B_{44}-B_{11}B_{22}B_{34}B_{43}-B_{11}B_{32}B_{23}B_{44}+B_{11}B_{42}B_{23}B_{34}
\nonumber  \\
&-&B_{21}B_{12}B_{33}B_{44}+B_{21}B_{12}B_{34}B_{43}+B_{21}B_{32}B_{13}B_{44}-B_{21}B_{42}B_{13}B_{34}.
\end{eqnarray}

So, the Mie coefficient $b^{\rm r}_{n}$ is of the form

\begin{equation}
b^{\rm r}_{n}=\frac{{\rm det} B_{\rm r}}{{\rm det} B}
\end{equation}
with
\begin{equation}
B_{\rm r}=\left(\begin{array}{cccc}
B_{11} & B_{12} & B_{13}& 0 \\
B_{21} & B_{22}& B_{23}& 0  \\
 0 & B_{32} & B_{33} &  N_{2}\left[\rho_{2}j_{n}(\rho_{2})\right]' \\
 0&B_{42}&B_{43}&\mu_{2}j_{n}(\rho_{2})
 \end{array}
 \right)
\end{equation}
whose determinant is

\begin{eqnarray}
{\rm det} B_{\rm r
}&=&B_{11}B_{22}B_{33}\mu_{2}j_{n}(\rho_{2})-B_{11}B_{22}N_{2}\left[\rho_{2}j_{n}(\rho_{2})\right]'B_{43}-B_{11}B_{32}B_{23}\mu_{2}j_{n}(\rho_{2})+B_{11}B_{42}B_{23}N_{2}\left[\rho_{2}j_{n}(\rho_{2})\right]'
 \nonumber   \\
&-&B_{21}B_{12}B_{33}\mu_{2}j_{n}(\rho_{2})+B_{21}B_{12}N_{2}\left[\rho_{2}j_{n}(\rho_{2})\right]'B_{43}+B_{21}B_{32}B_{13}\mu_{2}j_{n}(\rho_{2})-B_{21}B_{42}B_{13}N_{2}\left[\rho_{2}j_{n}(\rho_{2})\right]'.
\end{eqnarray}
\section{Reduced to the case of single-layered sphere}
In order to see whether the above Mie coefficients in the case of
double-layered sphere is correct or not, we consider the reduction
problem of double-layered case to the single-layered one when the
following reduction conditions are satisfied: $A_{11}=-A_{12}$,
$A_{21}=-A_{22}$ $N_{1}=N_{2}$, $\rho_{1}=\rho_{2}$,
$\mu_{1}=\mu_{2}$.

By lengthy calculation, we obtain
\begin{eqnarray}
{\rm
det}A&=&\left(A_{11}A_{23}-A_{21}A_{13}\right)\left(A_{42}A_{34}-A_{32}A_{44}\right)
\nonumber \\
&=&\left(A_{11}A_{23}-A_{21}A_{13}\right)\left\{j_{n}(N_{2}\rho_{2})\mu_{2}\left[\rho_{2}h^{(1)}_{n}(\rho_{2})\right]'-\mu_{0}\left[N_{2}\rho_{2}j_{n}(N_{2}\rho_{2})\right]'h^{(1)}_{n}(\rho_{2})\right\}
\end{eqnarray}
and
\begin{equation}
{\rm det}A_{\rm
r}=\left(A_{11}A_{23}-A_{21}A_{13}\right)\left\{\mu_{0}\left[N_{2}\rho_{2}j_{n}(N_{2}\rho_{2})\right]'j_{n}(\rho_{2})-j_{n}(N_{2}\rho_{2})\mu_{2}\left[\rho_{2}j_{n}(\rho_{2})\right]'\right\}.
\end{equation}
Thus
\begin{equation}
a^{\rm
r}_{n}=\frac{\mu_{0}\left[N_{2}\rho_{2}j_{n}(N_{2}\rho_{2})\right]'j_{n}(\rho_{2})-j_{n}(N_{2}\rho_{2})\mu_{2}\left[\rho_{2}j_{n}(\rho_{2})\right]'}{j_{n}(N_{2}\rho_{2})\mu_{2}\left[\rho_{2}h^{(1)}_{n}(\rho_{2})\right]'-\mu_{0}\left[N_{2}\rho_{2}j_{n}(N_{2}\rho_{2})\right]'h^{(1)}_{n}(\rho_{2})},
\end{equation}
which is just the Mie coefficient $a^{\rm r}_{n}$ of
single-layered sphere\cite{Stratton}.

 In the same fashion, when the reduction conditions $B_{11}=-B_{12}$, $B_{21}=-B_{22}$
$N_{1}=N_{2}$, $\rho_{1}=\rho_{2}$, $\mu_{1}=\mu_{2}$ are
satisfied, one can arrive at
\begin{eqnarray}
{\rm
det}B&=&\left(B_{11}B_{23}-B_{21}B_{13}\right)\left(B_{42}B_{34}-B_{32}B_{44}\right)
\nonumber \\
&=&\left(B_{11}B_{23}-B_{21}B_{13}\right)\left\{\left[N_{2}\rho_{2}j_{n}(N_{2}\rho_{2})\right]'\mu_{2}h^{(1)}_{n}(\rho_{2})-N^{2}_{2}\mu_{0}j_{n}(N_{2}\rho_{2})\left[\rho_{2}h^{(1)}_{n}(\rho_{2})\right]'\right\}
\end{eqnarray}
and
\begin{equation}
{\rm det}B_{\rm
r}=\left(B_{11}B_{23}-B_{21}B_{13}\right)\left\{N^{2}_{2}\mu_{0}j_{n}(N_{2}\rho_{2})\left[\rho_{2}j_{n}(\rho_{2})\right]'-\left[N_{2}\rho_{2}j_{n}(N_{2}\rho_{2})\right]'\mu_{2}j_{n}(\rho_{2})\right\}.
\end{equation}
So,
\begin{equation}
b^{\rm
r}_{n}=\frac{N^{2}_{2}\mu_{0}j_{n}(N_{2}\rho_{2})\left[\rho_{2}j_{n}(\rho_{2})\right]'-\left[N_{2}\rho_{2}j_{n}(N_{2}\rho_{2})\right]'\mu_{2}j_{n}(\rho_{2})}{\left[N_{2}\rho_{2}j_{n}(N_{2}\rho_{2})\right]'\mu_{2}h^{(1)}_{n}(\rho_{2})-N^{2}_{2}\mu_{0}j_{n}(N_{2}\rho_{2})\left[\rho_{2}h^{(1)}_{n}(\rho_{2})\right]'},
\end{equation}
which is just the Mie coefficient $b^{\rm r}_{n}$ of
single-layered sphere\cite{Stratton}.

This, therefore, means that the Mie coefficients of double-layered
sphere presented here is right.
\section{All the Mie coefficients inside the double-layered sphere}
\subsection{Mie coefficients $a^{\rm t}_{n}$, $a^{\rm m}_{n}$ and $a^{\bar{\rm m}}_{n}$}
The $4\times4$ matrix $A_{\rm t}$ is
\begin{equation}
A_{\rm t}=\left(\begin{array}{cccc}
0 & A_{12} & A_{13}& 0 \\
0 & A_{22}& A_{23}& 0  \\
  j_{n}(\rho_{2})& A_{32} & A_{33} & A_{34}\\
 \mu_{2}\left[\rho_{2}j_{n}(\rho_{2})\right]'&A_{42}&A_{43}&A_{44}
 \end{array}
 \right)
\end{equation}
whose determinant is
\begin{equation}
{\rm det}A_{\rm
t}=j_{n}(\rho_{2})A_{12}A_{23}A_{44}-j_{n}(\rho_{2})A_{22}A_{13}A_{44}-\mu_{2}\left[\rho_{2}j_{n}(\rho_{2})\right]'A_{12}A_{23}A_{34}+\mu_{2}\left[\rho_{2}j_{n}(\rho_{2})\right]'A_{22}A_{13}A_{34}.
\end{equation}
So,
\begin{equation}
a^{\rm t}_{n}=\frac{{\rm det} A_{\rm t}}{{\rm det} A}.
\end{equation}

The $4\times4$ matrix $A_{\rm m}$ is
\begin{equation}
A_{\rm m}=\left(\begin{array}{cccc}
A_{11} & 0& A_{13}& 0 \\
A_{21} & 0& A_{23}& 0  \\
  0 & j_{n}(\rho_{2}) & A_{33} & A_{34}\\
 0 & \mu_{2}\left[\rho_{2}j_{n}(\rho_{2})\right]'& A_{43}& A_{44}
 \end{array}
 \right)
\end{equation}
whose determinant is
\begin{equation}
{\rm det}A_{\rm
m}=-A_{11}j_{n}(\rho_{2})A_{23}A_{44}+A_{11}\mu_{2}\left[\rho_{2}j_{n}(\rho_{2})\right]'A_{23}A_{34}+A_{21}j_{n}(\rho_{2})A_{13}A_{44}-A_{21}\mu_{2}\left[\rho_{2}j_{n}(\rho_{2})\right]'A_{13}A_{34}.
\end{equation}
So,
\begin{equation}
a^{\rm m}_{n}=\frac{{\rm det} A_{\rm m}}{{\rm det} A}.
\end{equation}

The $4\times4$ matrix $A_{\bar{\rm m}}$ is
\begin{equation}
A_{\bar{\rm m}}=\left(\begin{array}{cccc}
A_{11} & A_{12}& 0& 0 \\
A_{21} & A_{22}& 0& 0  \\
  0 & A_{32} & j_{n}(\rho_{2})  & A_{34}\\
 0 & A_{42}& \mu_{2}\left[\rho_{2}j_{n}(\rho_{2})\right]'& A_{44}
 \end{array}
 \right)
\end{equation}
whose determinant is
\begin{equation}
{\rm det}A_{\bar{\rm m}}=A_{11}A_{22}j_{n}(\rho_{2})
A_{44}-A_{11}A_{22}A_{34}\mu_{2}\left[\rho_{2}j_{n}(\rho_{2})\right]'-A_{21}A_{12}j_{n}(\rho_{2})A_{44}+A_{21}A_{12}A_{34}\mu_{2}\left[\rho_{2}j_{n}(\rho_{2})\right]'.
\end{equation}
So,
\begin{equation}
a^{\bar{\rm m}}_{n}=\frac{{\rm det}A_{\bar{\rm m}}}{{\rm det} A}.
\end{equation}
\\ \\

It is readily verified that under the reduction conditions
$A_{11}=-A_{12}$, $A_{21}=-A_{22}$, $N_{1}=N_{2}$,
$\rho_{1}=\rho_{2}$, $\mu_{1}=\mu_{2}$, one can arrive at
\begin{equation}
{\rm det}A_{\rm t}={\rm det}A_{\rm m}, \quad a^{\rm t}_{n}=a^{\rm
m}_{n},  \quad {\rm det}A_{\bar{\rm m}}=0,     \quad   a^{\bar{\rm
m}}_{n}=0,
\end{equation}
which means that the above Mie coefficient can be reduced to the
those of single-layered sphere.
\subsection{Mie coefficients $b^{\rm t}_{n}$, $b^{\rm m}_{n}$ and $b^{\bar{\rm m}}_{n}$}

The $4\times4$ matrix $B_{\rm t}$ is
\begin{equation}
B_{\rm t}=\left(\begin{array}{cccc}
0 & B_{12} & B_{13}& 0 \\
0 & B_{22}& B_{23}& 0  \\
  N_{2}\left[\rho_{2}j_{n}(\rho_{2})\right]' & B_{32} & B_{33} & B_{34}\\
 \mu_{2}j_{n}(\rho_{2})&B_{42}&B_{43}&B_{44}
 \end{array}
 \right)
\end{equation}
whose determinant is
\begin{equation}
{\rm det}B_{\rm t}=N_{2}\left[\rho_{2}j_{n}(\rho_{2})\right]'
B_{12}B_{23}B_{44}-N_{2}\left[\rho_{2}j_{n}(\rho_{2})\right]'
B_{22}B_{13}B_{44}-\mu_{2}j_{n}(\rho_{2})
B_{12}B_{23}B_{34}+\mu_{2}j_{n}(\rho_{2})B_{22}B_{13}B_{34}.
\end{equation}
So,
\begin{equation}
b^{\rm t}_{n}=\frac{{\rm det} B_{\rm t}}{{\rm det} B}.
\end{equation}
\\ \\
The $4\times4$ matrix $B_{\rm m}$ is
\begin{equation}
B_{\rm m}=\left(\begin{array}{cccc}
B_{11} & 0& B_{13}& 0 \\
B_{21} & 0& B_{23}& 0  \\
  0 & N_{2}\left[\rho_{2}j_{n}(\rho_{2})\right]'  & B_{33} & B_{34}\\
 0 & \mu_{2}j_{n}(\rho_{2})& B_{43}& B_{44}
 \end{array}
 \right)
\end{equation}
whose determinant is
\begin{equation}
{\rm det}B_{\rm
m}=-B_{11}N_{2}\left[\rho_{2}j_{n}(\rho_{2})\right]'
B_{23}B_{44}+B_{11}\mu_{2}j_{n}(\rho_{2})
B_{23}B_{34}+B_{21}N_{2}\left[\rho_{2}j_{n}(\rho_{2})\right]'
B_{13}B_{44}-B_{21}\mu_{2}j_{n}(\rho_{2})B_{13}B_{34}.
\end{equation}
So,
\begin{equation}
b^{\rm m}_{n}=\frac{{\rm det} B_{\rm m}}{{\rm det} B}.
\end{equation}
\\ \\
The $4\times4$ matrix $B_{\bar{\rm m}}$ is
\begin{equation}
B_{\bar{\rm m}}=\left(\begin{array}{cccc}
B_{11} & B_{12}& 0& 0 \\
B_{21} & B_{22}& 0& 0  \\
  0 & B_{32} & N_{2}\left[\rho_{2}j_{n}(\rho_{2})\right]'   & B_{34}\\
 0 & B_{42}& \mu_{2}j_{n}(\rho_{2})& B_{44}
 \end{array}
 \right)
\end{equation}
whose determinant is
\begin{equation}
{\rm det}B_{\bar{\rm
m}}=B_{11}B_{22}N_{2}\left[\rho_{2}j_{n}(\rho_{2})\right]'
B_{44}-B_{11}B_{22}B_{34}\mu_{2}j_{n}(\rho_{2})
-B_{21}B_{12}N_{2}\left[\rho_{2}j_{n}(\rho_{2})\right]'
B_{44}+B_{21}B_{12}B_{34}\mu_{2}j_{n}(\rho_{2}).
\end{equation}
So,
\begin{equation}
b^{\bar{\rm m}}_{n}=\frac{{\rm det}B_{\bar{\rm m}}}{{\rm det} B}.
\end{equation}
\\ \\

It is readily verified that under the reduction conditions
$B_{11}=-B_{12}$, $B_{21}=-B_{22}$,   $N_{1}=N_{2}$,
$\rho_{1}=\rho_{2}$, $\mu_{1}=\mu_{2}$, one can arrive at
\begin{equation}
{\rm det}B_{\rm t}={\rm det}B_{\rm m}, \quad b^{\rm t}_{n}=b^{\rm
m}_{n},  \quad {\rm det}B_{\bar{\rm m}}=0,     \quad   b^{\bar{\rm
m}}_{n}=0,
\end{equation}
which means that the above Mie coefficient can be reduced to the
those of single-layered sphere.

\section{Discussion of some typical cases with left-handed media involved}
In this section we briefly discuss several cases with left-handed
media involved.

(i) If the following conditions $h^{(1)}_{n}(N_{2}\rho_{1})=0$,
$\left[h^{(1)}_{n}(N_{2}\rho_{1})\right]'=0$,
$h^{(1)}_{n}(N_{2}\rho_{2})=-h^{(1)}_{n}(\rho_{2})$,
$\left[h^{(1)}_{n}(N_{2}\rho_{2})\right]'=-\left[h^{(1)}_{n}(\rho_{2})\right]'$,
$N_{2}=\mu_{2}=\epsilon_{2}=-1$ and $\mu_{0}=+1$ are satisfied,
then it is easily verified that

\begin{equation}
{\rm det}A=0,   \quad     {\rm det}B=0     \quad  {\rm and} \quad
a^{\rm r}_{n}=\infty,    \quad     b^{\rm r}_{n}=\infty.
\end{equation}
Thus in this case the extinction cross section of the two-layered
sphere containing left-handed media is rather large.

(ii) If the following conditions $h^{(1)}_{n}(N_{2}\rho_{1})=0$,
$\left[h^{(1)}_{n}(N_{2}\rho_{1})\right]'=0$,
$h^{(1)}_{n}(N_{2}\rho_{2})=j_{n}(\rho_{2})$,
$\left[h^{(1)}_{n}(N_{2}\rho_{2})\right]'=\left[j_{n}(\rho_{2})\right]'$,
$N_{2}=\mu_{2}=\epsilon_{2}=-1$ and $\mu_{0}=+1$ are satisfied,
then it is easily verified that

\begin{equation}
{\rm det}A_{\rm r}=0,   \quad     {\rm det}B_{\rm r}=0     \quad
{\rm and} \quad a^{\rm r}_{n}=0,    \quad     b^{\rm r}_{n}=0.
\end{equation}
Thus in this case the extinction cross section of the two-layered
sphere containing left-handed media is negligibly small.

(iii) The case with $N_{1}=\mu_{1}=\epsilon_{1}=-1$,
$N_{2}=\mu_{2}=\epsilon_{2}=0$, $N_{0}=\mu_{0}=\epsilon_{0}=+1$ is
of physical interest, which deserves consideration by using the
Mie coefficients presented above.

Based on the calculation of Mie coefficients, one can treat the
scattering problem of double-layered sphere irradiated by a plane
wave. The scattering and absorption properties of double-layered
sphere containing left-handed media can thus be discussed in
detail, which are now under consideration and will be submitted
elsewhere for the publication.
\\ \\

\textbf{Acknowledgements}   This project is supported by the
National Natural Science Foundation of China under the project No.
$90101024$.


\begin{references}
\bibitem{Smith} Smith, D. R., Padilla, W. J., Vier, D. C. {\it et
al.}, Phys. Rev. Lett. {\bf 84}, 4184 (2000).

\bibitem{Klimov} Klimov, V. V., Opt. Comm. {\bf 211}, 183 (2002).

\bibitem{Pendry3}  Pendry, J. B., Holden, A. J., Robbins, D. J. and
Stewart, W. J., IEEE Trans. Microwave Theory Tech. {\bf 47}, 2075
(1999).

\bibitem{Shelby} Shelby, R. A., Smith, D. R. and Schultz, S.,
Science {\bf 292}, 77 (2001).

\bibitem{Ziolkowski2} Ziolkowski, R. W., Phys. Rev. E {\bf 64},
056625 (2001).

\bibitem{Veselago} Veselago, V. G., Sov. Phys. Usp. {\bf 10}, 509
(1968).

\bibitem{Pendry1} Pendry, J. B., Holden, A. J., Stewart, W. J. and
Youngs, I., Phys. Rev. Lett. {\bf 76}, 4773 (1996).

\bibitem{Pendry2} Pendry, J. B., Holden, Robbins, D. J. and
Stewart, W. J., J. Phys. Condens. Matter {\bf 10}, 4785 (1998).

\bibitem{Maslovski} Maslovski, S. I., Tretyakov, S. A. and Belov,
P. A., Inc. Microwave Opt. Tech. Lett. {\bf 35}, 47 (2002).

\bibitem{Wang} Wang, W. and Asher, S. A., J. Am. Chem. Soc. {\bf
123}, 12528 (2001).

\bibitem{Stratton}  Stratton, J. A., Electromagnetic theory (New York:
McGraw-Hill) (1941).
\end{references}
\end{document}